\documentclass[letterpaper,journal]{IEEEtran}
\usepackage{amsmath,amsfonts}
\usepackage{amssymb}
\usepackage{algorithm}
\usepackage{algpseudocode}
\usepackage{array}
\usepackage[caption=false,font=normalsize,labelfont=sf,textfont=sf]{subfig}
\usepackage{textcomp}
\usepackage{stfloats}
\usepackage{multirow}
\usepackage{booktabs}
\usepackage{url}
\usepackage{verbatim}
\usepackage{graphicx}
\usepackage{xcolor}
\usepackage{cite}
\begin{document}

\title{Generative Video Compression with Adaptive Score Distillation}

\author{Naifu~Xue, Zhaoyang~Jia, Haosen~Li, Zihan~Zheng, \\Jiahao~Li, Bin~Li, Xiaoyi~Zhang, Qi~Meng, Yuan~Zhang, and~Yan~Lu%
\thanks{N. Xue and Y. Zhang are with the Communication University of China, Beijing, China (e-mail: \{aaronxuenf, yzhang\}@cuc.edu.cn).}%
\thanks{Z. Jia and Z. Zheng are with the University of Science and Technology of China, Hefei, China (e-mail: \{jzy\_ustc, zzh2003\}@mail.ustc.edu.cn).}%
\thanks{J. Li, B. Li, X. Zhang, and Y. Lu are with Microsoft Research Asia (e-mail: \{li.jiahao, libin, xiaoyizhang, yanlu\}@microsoft.com).}%
\thanks{H. Li and Q. Meng are with Academy of Mathematics and Systems Science, Chinese Academy of Sciences, Beijing, China (e-mail: lihaosen25@mails.ucas.ac.cn, meq@amss.ac.cn).}%
\thanks{N. Xue, Z. Jia, H. Li, and Z. Zheng are interns at Microsoft Research Asia.}}

\maketitle

\begin{abstract}
Diffusion models provide strong generative capabilities for video compression at ultra-low bitrates.
Existing diffusion-based video codecs adapt base models originally developed for text-conditioned generation, whereas diffusion models designed and trained specifically for compression remain unexplored.
To fill this gap, we introduce our \textbf{Generative Video Codec (GenVC)}, built on a video diffusion model trained from scratch for compression.
To our knowledge, this is the first compression-oriented video diffusion model.
We realize this model directly in pixel space with a global-to-local hierarchy that recovers fine spatio-temporal details, enabling high-quality generative reconstruction from compressed representations.
To accelerate inference, we distill the multi-step model into one step using distribution matching distillation (DMD).
Applying DMD directly, however, drives the student toward motion-stalled reconstructions. 
We trace this to a \emph{teacher-side guidance failure}: once student-induced perturbations leave the frozen teacher's training region, its guidance can become misleading, causing DMD updates to reinforce rather than correct the student drift.
To break the resulting feedback loop, we propose \emph{Adaptive Score Distillation}, which gates DMD updates according to their alignment with the ground-truth direction, enabling high-quality reconstruction with coherent motion.
Experimental results show that GenVC achieves state-of-the-art perceptual quality at ultra-low bitrates, with average bitrate savings of 62.5\% at matched LPIPS and 71.3\% at matched FID over GLVC.
Unlike prior codecs that inherit billion-scale pretrained backbones, our diffusion model has only 478.0M parameters and decodes 1080p video in a single step at 15.1 fps on an A100 GPU.
\end{abstract}

\begin{IEEEkeywords}
Neural video compression, diffusion model, distribution matching distillation, one-step diffusion.
\end{IEEEkeywords}

\section{Introduction}

\begin{figure}[t]
    \centering
    \includegraphics[width=1.0\linewidth]{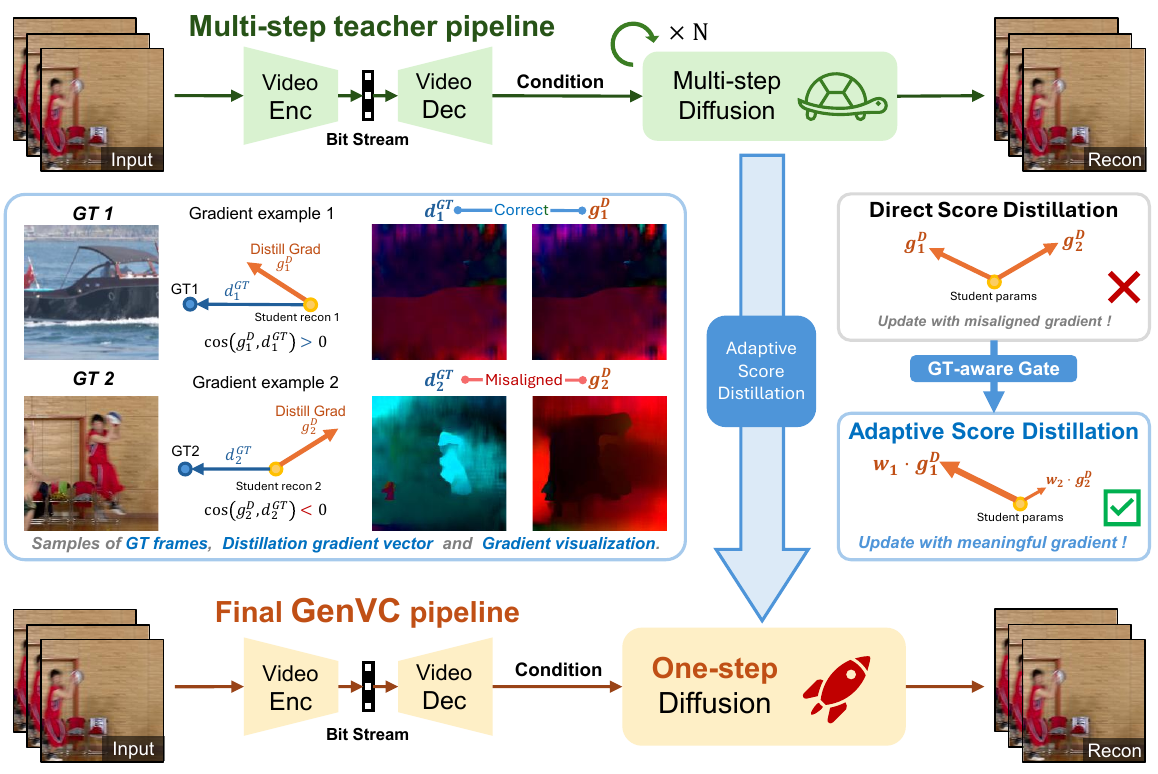}
    \vspace{-7mm}
    \caption{
        Overview of GenVC. 
        The multi-step model (top) serves as the teacher for one-step \textbf{GenVC} (bottom).
        The proposed \emph{Adaptive Score Distillation} gates DMD updates according to their alignment with the ground-truth direction, downweighting poorly aligned updates while preserving informative ones.
    }
    \label{fig:1}
    \vspace{-6mm}
\end{figure}

\IEEEPARstart{G}{enerative} video compression has emerged to improve perceptual quality at ultra-low bitrates~\cite{yang2022perceptual,ma2025diffusion,qi2025generative}. By integrating generative models into the codec, these methods synthesize plausible details that are not explicitly transmitted in the bitstream.
Among these generative models, native video diffusion~\cite{zheng2024opensora,yang2025cogvideox,kong2024hunyuanvideo,wan2025wan} has recently attracted growing attention for its strong spatio-temporal modeling capacity, making it an appealing choice for generative video codecs. 
Existing diffusion-based video codecs~\cite{mao2026generative,zeng2026gvcc,zhou2026efficient} commonly adapt pretrained latent diffusion backbones developed for text-conditioned generation.
While effective, this leaves open whether a video diffusion model can instead be learned jointly with the codec, allowing the compression objective to shape both the transmitted representation and the diffusion model.
On the image side, compression-oriented diffusion~\cite{Jia_2026_CVPR,jia2026codliterealtimediffusionbasedgenerative} provides an encouraging precedent: under strong codec conditioning, a pixel-domain diffusion model can be trained from scratch jointly with the codec using only a fraction of the data and computation used for large-scale text-to-image pretraining, while achieving state-of-the-art (SOTA) compression performance.
Inspired by this precedent, we introduce our \textbf{Generative Video Codec (GenVC)}, built on a video diffusion model jointly optimized with the codec.
To our knowledge, this is the first compression-oriented video diffusion model trained from scratch.
The resulting multi-step model achieves strong reconstruction performance on its own, demonstrating effective generative modeling tailored to video compression.

Achieving this performance directly in pixel space, however, requires addressing a key architectural challenge: a plain DiT reconstructs full-resolution video from a coarse spatio-temporal token grid. 
While this coarse grid keeps global modeling tractable, direct unpatchification tends to miss fine spatial texture and local temporal variation. 
We therefore develop a global-to-local refinement hierarchy: a coarse spatio-temporal DiT captures long-range structure, followed by patch-level and frame-level refiners that progressively recover fine spatial and temporal detail for high-quality reconstruction.

Despite its strong performance, the multi-step model relies on iterative sampling, making decoding slow.
We therefore use it as the teacher for one-step GenVC, as shown in Fig.~\ref{fig:1}.
One-step distillation is a natural acceleration strategy and an active research direction~\cite{ma2025diffvc,ma2026diffvc,xue2026single}.
For this purpose, distribution matching distillation (DMD)~\cite{Yin_2024_CVPR,yin2024improved} provides a mature score-distillation approach whose effectiveness has been demonstrated in image generation and compression~\cite{NEURIPS2025_352a67c0,jia2026codliterealtimediffusionbasedgenerative}.
We therefore adopt DMD to distill our multi-step teacher into the one-step GenVC.
However, directly applying DMD in this setting drives the student toward motion-stalled reconstructions.
We trace this behavior to a \emph{teacher-side guidance failure}.
In our design, the frozen multi-step teacher serves as DMD's real score network.
Once motion stalling emerges in the student reconstructions, their perturbations move beyond the region covered during teacher training.
On such inputs, the teacher prediction can fail to restore the missing motion and instead suppress it further.
The resulting DMD update can then push the student farther from the source, producing increasingly off-distribution inputs to the teacher and forming a self-reinforcing degradation loop.

\begin{figure*}[t]
    \centering
    \includegraphics[width=1.0\linewidth]{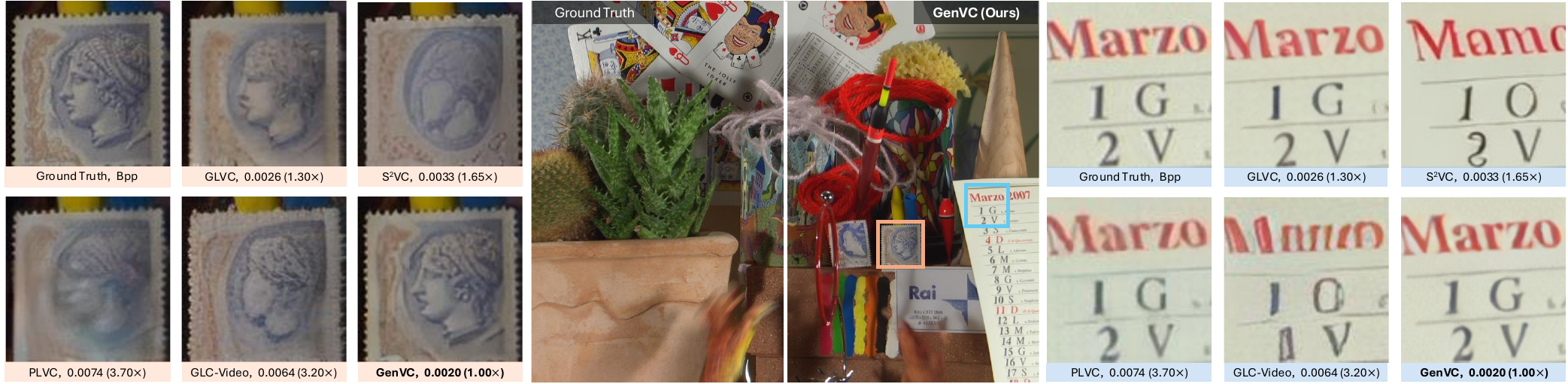}
    \vspace{-7mm}
    \caption{
        Frame examples. 
        GenVC delivers the best perceptual quality at the lowest bitrate among the compared methods. Zoom in for better visualization.
    }
    \label{fig:2}
    \vspace{-4mm}
\end{figure*}

\begin{figure}[t]
    \centering
    \includegraphics[width=1.0\linewidth]{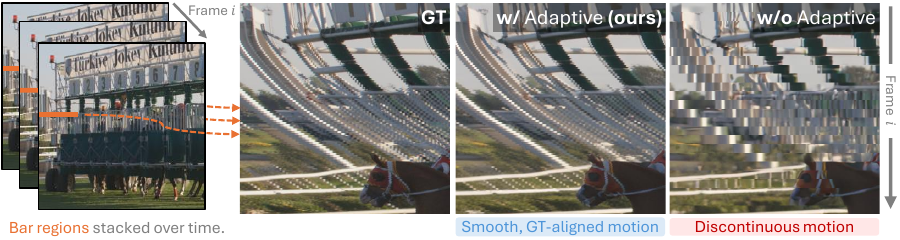}
    \vspace{-7mm}
    \caption{
        \textbf{Adaptive gating prevents motion stalling.}
        Horizontal strips extracted from consecutive frames are stacked from top to bottom, illustrating the motion trajectory over time.
        Close agreement with the GT strip indicates faithful motion reconstruction, whereas pronounced staircase patterns reveal motion stalling and abrupt jumps between adjacent frames.
    }
    \label{fig:2_1}
    \vspace{-5mm}
\end{figure}

Concretely, we find that the alignment between the DMD update and the ground-truth direction indicates whether the update becomes misleading (Fig.~\ref{fig:1}, middle).
Building on this observation, we introduce \textbf{Adaptive Score Distillation}, which downweights misleading DMD updates while preserving informative ones.
Together with the ground-truth-anchored distortion losses already used in codec training (e.g., $\ell_1$ and LPIPS~\cite{zhang2018unreasonable}), this selective gating stabilizes distillation while retaining the realism supplied by DMD.
Fig.~\ref{fig:2_1} visualizes this stabilization: gated distillation preserves ground-truth-aligned motion, whereas removing the gate produces staircase patterns that reveal motion stalling.

Experimental results show that GenVC achieves leading perceptual quality at ultra-low bitrates, with average bitrate savings of 62.5\% at matched LPIPS and 71.3\% at matched FID over GLVC~\cite{guo2025generative}.
Fig.~\ref{fig:2} corroborates this qualitatively: at the lowest bitrate among the compared codecs, GenVC reconstructs higher-fidelity detail while excelling in realism.
The resulting diffusion model has 478.0M parameters, demonstrating that compression-oriented video diffusion can be realized at such scale without inheriting a billion-scale pretrained backbone like in existing solutions.
Distilling it to one step raises 1080p decoding throughput from 0.40 fps for the multi-step teacher to 15.1 fps for GenVC on an A100 GPU.

In summary, our contributions are as follows:
\begin{itemize}
    \item We introduce the first video diffusion codec whose diffusion model is trained from scratch, and develop a global-to-local hierarchy that enables strong standalone multi-step reconstruction in pixel space.
    \item We identify a \emph{teacher-side guidance failure} in distillation: off-distribution student-induced inputs can make the frozen teacher produce misleading DMD updates that reinforce motion stalling.
    \item We propose \emph{Adaptive Score Distillation}, which filters poorly aligned DMD updates to stabilize one-step distillation in ground-truth-anchored reconstruction.
    \item GenVC achieves SOTA perceptual quality with average bitrate savings exceeding 60\% on both LPIPS and FID relative to GLVC, using a 478.0M diffusion model and reaching 15.1 fps for 1080p decoding on an A100 GPU.
\end{itemize}

\section{Related Work}
\label{sec:related}

\subsection{Generative Video Compression}
\label{sec:rw-generative}

Neural video codecs~\cite{li2021deep, li2023neural, li2024neural, jia2025towards, li2026ultra} have surpassed traditional standards~\cite{bross2021overview} on objective metrics, yet at low bitrates their reconstructions become over-smoothed because the limited rate budget is insufficient to represent fine texture.
To improve perceptual quality at such bitrates, generative video codecs synthesize plausible details that are not explicitly transmitted in the bitstream.
Early methods train decoders with perceptual and adversarial objectives to restore realistic texture~\cite{mentzer2022neural,yang2022perceptual}.
More recent approaches instead perform coding in the latent space of a generative tokenizer, improving compression efficiency and reconstruction quality~\cite{qi2025generative,guo2025generative}.
However, their limited generative capacity constrains perceptual quality at ultra-low bitrates.

Diffusion models have recently become prominent in generative video compression because of their impressive visual synthesis capacity.
Most diffusion-based video codecs adapt billion-scale pretrained backbones, such as Stable Diffusion~\cite{Rombach_2022_CVPR} and Wan~\cite{wan2025wan}, to reconstruct frames~\cite{ma2025diffusion,mao2026generative,zhou2026efficient} with multi-step inference.
To accelerate decoding, recent methods adopt one-step diffusion~\cite{ma2025diffvc,ma2026diffvc,xue2026single,li2026yoda}, yet their large pretrained backbones continue to limit decoding throughput.
GenVC instead jointly learns a compression-oriented diffusion model with its codec, focusing its modeling capacity on compression-specific detail synthesis.
This specialization allows GenVC to achieve SOTA perceptual quality at ultra-low bitrates with a sub-billion-parameter diffusion model.
Its distilled one-step model also delivers substantially faster inference than existing diffusion-based video codecs.

\subsection{Pixel-Domain Diffusion}
\label{sec:rw-pixel}

On the image side, recent studies show that pixel-domain diffusion can achieve competitive generation quality~\cite{Li_2026_CVPR,wang2026pixnerd,Chen_2026_CVPR_DiP,Ma_2026_CVPR_DeCo,Yu_2026_CVPR_PixelDiT}.
This formulation provides a simplified single-stage training pipeline~\cite{Li_2026_CVPR,wang2026pixnerd} and avoids the fidelity loss introduced by lossy VAE reconstruction~\cite{wang2026pixnerd,Chen_2026_CVPR_DiP,Ma_2026_CVPR_DeCo,Yu_2026_CVPR_PixelDiT}, highlighting its potential for high-fidelity generation.
To make full-resolution pixel modeling tractable, recent methods adopt coarse-to-fine architectures that pair a transformer for global structure with lightweight refiners for local detail~\cite{wang2026pixnerd,Chen_2026_CVPR_DiP,Ma_2026_CVPR_DeCo,Yu_2026_CVPR_PixelDiT}.
The CoD series further applies this paradigm to image compression, achieving SOTA performance~\cite{Jia_2026_CVPR,jia2026codliterealtimediffusionbasedgenerative} without pretrained backbones.

Yet, pixel-domain diffusion remains largely unexplored for video due to the higher spatio-temporal complexity of video data.
One-DVA~\cite{teng2026onedva} explores pixel diffusion within a video autoencoder but focuses on low-resolution reconstruction.
Meanwhile, although existing diffusion-based video codecs~\cite{mao2026generative,ma2025diffvc,ma2026diffvc,xue2026single,li2026yoda} perform diffusion in latent space, they still impose reconstruction objectives on decoded pixels to preserve source fidelity.
Together, these observations motivate GenVC to learn diffusion directly in pixel space.
We make this approach practical for high-resolution video compression through an efficient global-to-local hierarchy.

\begin{figure*}[t]
    \centering
    \includegraphics[width=1.0\linewidth]{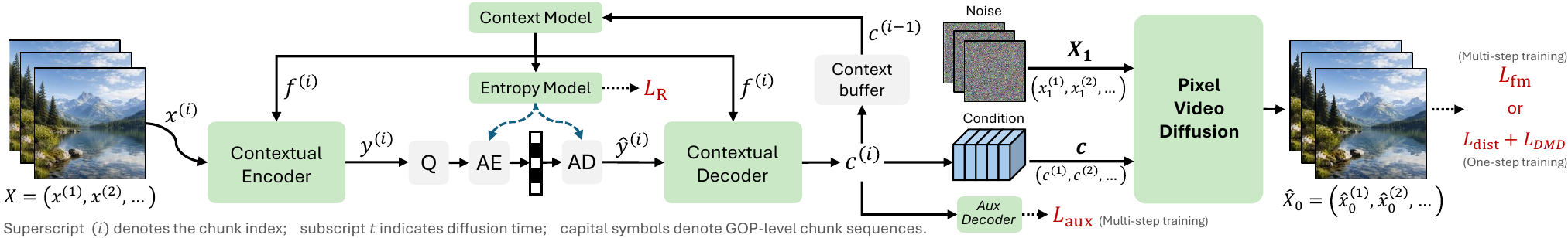}
    \vspace{-7mm}
    \caption{
        Overview of the \textbf{GenVC} pipeline.
        The clean source GOP and its decoded condition sequence are denoted by $X_0\equiv X=(x^{(1)},\ldots,x^{(K)})$ and $c=(c^{(1)},\ldots,c^{(K)})$, respectively.
        Q denotes quantization; AE and AD denote the arithmetic encoder and decoder.
    }
    \label{fig:3_1}
    \vspace{-4mm}
\end{figure*}

\subsection{Diffusion Acceleration and Score Distillation}
\label{sec:rw-distillation}

Reducing a multi-step diffusion teacher to a few- or one-step student is a widely studied topic.
Score-distillation methods~\cite{alldieck2024score,wang2023prolificdreamer} optimize differentiable representations or generators through a frozen diffusion prior.
Building on this idea, distribution matching distillation (DMD)~\cite{Yin_2024_CVPR,yin2024improved} matches the student's and teacher's output distributions to enable powerful few- and one-step distillation.
DMD and its variants have since been widely adopted in one-step diffusion codecs~\cite{NEURIPS2025_352a67c0,Jia_2026_CVPR,jia2026codliterealtimediffusionbasedgenerative} and extended to few-step video generation~\cite{yin2025slow}.
Building on this line, our Adaptive Score Distillation equips DMD with an adaptive gate that stabilizes video codec distillation and enables high-quality one-step reconstruction.

\section{Compression-Oriented Video Diffusion}
\label{sec:pixvdcm}

We first develop the compression-oriented video diffusion model underlying GenVC.
Rather than adapting a generative backbone pretrained for another task, we train a pixel-domain diffusion model from scratch jointly with a contextual video codec (Fig.~\ref{fig:3_1}).
This coupling allows the transmitted representation and conditional diffusion model to co-adapt for rate-constrained reconstruction, thereby improving reconstruction quality at a given bitrate.

Following~\cite{li2026ultra}, we partition each video into consecutive 8-frame chunks, each coded as one unit.
For chunk $x^{(i)}$, the contextual encoder produces a compact representation $y^{(i)}$.
After quantization, a learned entropy model provides probability estimates for arithmetic coding, and the contextual decoder maps the decoded representation $\hat{y}^{(i)}$ to the condition $c^{(i)}$.
Previously decoded conditions provide causal temporal context throughout this coding process, thereby exploiting inter-chunk redundancy to reduce rate.
At the diffusion stage, we group $K$ consecutive chunks into a group of pictures (GOP).
The remainder of this section reviews the flow-matching formulation (Sec.~\ref{sec:pixvdcm-bg}), presents the pixel diffusion architecture (Sec.~\ref{sec:pixvdcm-arch}), and details the training objective (Sec.~\ref{sec:pixvdcm-train}).

\begin{figure}[t]
    \centering
    \includegraphics[width=1.0\linewidth]{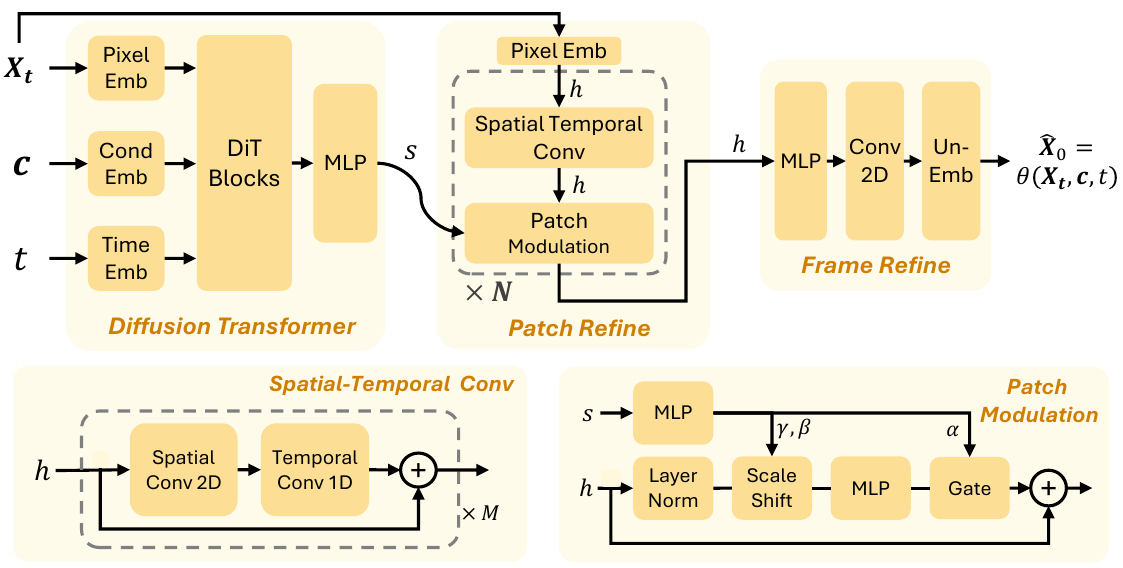}
    \vspace{-7mm}
    \caption{
        Global-to-local pixel-space architecture.
        A coarse spatio-temporal DiT captures long-range structure, Patch Refine blocks recover local spatial and temporal variation on a finer feature grid, and the Frame Refine head maps the refined features to RGB pixels.
    }
    \label{fig:3_2}
    \vspace{-4mm}
\end{figure}

\subsection{Flow Matching Preliminaries}
\label{sec:pixvdcm-bg}

Flow matching~\cite{lipman2023flow, liu2023flow} learns a time-dependent velocity field associated with a prescribed probability path between the data and noise distributions.
Using the data-to-noise time convention, let the source GOP $X\sim p_{\text{data}}$ define the clean diffusion state $X_0\equiv X$ at $t=0$, and let $X_1 \sim \mathcal{N}(\mathbf{0}, \mathbf{I})$ denote the independent Gaussian noise state at $t=1$.
Here, $t\in[0,1]$ denotes the diffusion time rather than a video-frame index.
We define the linear rectified-flow path~\cite{liu2023flow} as
\begin{equation}
    X_t = (1-t)\, X_0 + t\, X_1, \qquad t \in [0, 1],
    \label{eq:fm-path}
\end{equation}
whose conditional velocity is $\mathrm{d}X_t/\mathrm{d}t = X_1 - X_0$.
As depicted in Fig.~\ref{fig:3_2}, our network predicts the clean GOP rather than the velocity directly.
From its output $\hat{X}_0 = \theta(X_t,c,t)$, we convert the clean prediction to the corresponding velocity as
\begin{equation}
    \hat{v}_\theta(X_t,c,t) = \frac{X_t - \hat{X}_0}{t}, \qquad t>0,
    \label{eq:fm-x2v}
\end{equation}
which is the exact conversion induced by Eq.~\eqref{eq:fm-path}; the training-time sampler excludes the clean endpoint $t=0$.
We then optimize the conditional flow-matching objective,
\begin{equation}
    L_{\mathrm{fm}}(\theta) = \mathbb{E}_{(X_0,c),\, t,\, X_1}
    \left[ \big\| \hat{v}_\theta(X_t,c,t) - (X_1 - X_0) \big\|_2^2 \right].
    \label{eq:fm-loss}
\end{equation}
At inference, we solve the learned ODE backward from $X_1$ with a multi-step solver to obtain the reconstructed GOP.
In our codec setting, $X_0$ is the ground-truth GOP and $c$ is the corresponding collection of decoded codec conditions.
This sampler defines the standalone multi-step codec; the same diffusion model subsequently provides the frozen real-score estimates used by DMD.

\subsection{Global-to-Local Pixel-Space Architecture}
\label{sec:pixvdcm-arch}

\begin{figure}[t]
    \centering
    \includegraphics[width=1.0\linewidth]{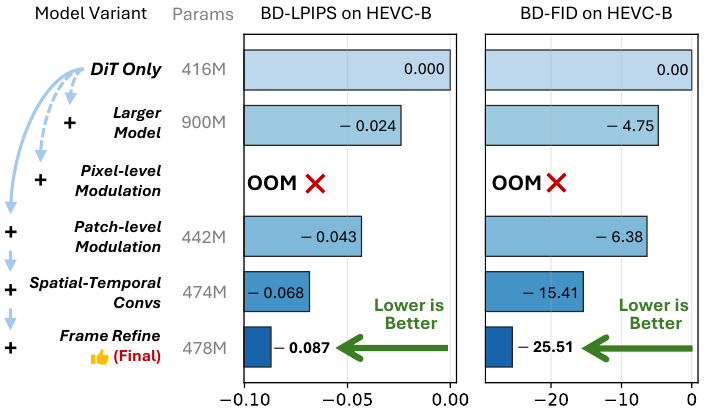}
    \vspace{-8mm}
    \caption{
        Progressive architecture ablation on HEVC-B, reported as BD-LPIPS and BD-FID relative to the DiT-only baseline.
        Under our 32-frame 1080p configuration, \emph{Pixel-level Modulation}, the standard per-pixel adaLN, exceeds the memory capacity of an 80\,GB GPU.
    }
    \label{fig:3_3}
    \vspace{-4mm}
\end{figure}

A natural starting point is to apply a plain pixel-space \emph{Diffusion Transformer} (DiT) directly to the stacked frames, following the image-side baseline~\cite{Li_2026_CVPR}.
To keep the pixel-space token sequence tractable, the DiT operates with a large patch size; in our setting, this baseline recovers coarse structure but fails to recover fine detail.
Simply increasing the DiT capacity yields only marginal perceptual gains (Fig.~\ref{fig:3_3}), indicating that model scaling alone is insufficient.
Reducing the spatial downsampling factor would provide a finer token grid but substantially lengthen the sequence.
Because self-attention scales quadratically with sequence length, diffusion transformers commonly operate at a spatial downsampling of $16\times$ or more~\cite{Li_2026_CVPR, wan2025wan}.
Following image-side pixel diffusion architectures~\cite{Li_2026_CVPR,wang2026pixnerd,Chen_2026_CVPR_DiP,Ma_2026_CVPR_DeCo,Yu_2026_CVPR_PixelDiT}, we adapt their global-to-local hierarchy to video.
The \emph{global} branch models long-range spatio-temporal structure, while the \emph{local} branch refines spatial and temporal detail on a finer feature grid.
In our design, the global branch embeds the noisy GOP $X_t$, the condition sequence $c$, and the diffusion time $t$, and processes them with DiT blocks on a token grid downsampled $16\times$ spatially (following common DiT practice) and $8\times$ temporally (following the codec design~\cite{li2026ultra}), yielding a patch-level global feature
\begin{equation}
    s = \mathrm{MLP}_{\text{g}}\!\left(\mathrm{DiT}(X_t,\, c,\, t)\right),
    \label{eq:arch-dit}
\end{equation}
which then guides the local branch through the \emph{Patch Refine} blocks and a final \emph{Frame Refine} head, as described next.

In the image-side designs we build on, the local branch runs at \emph{full pixel resolution}: the global feature $s$ is upsampled to the pixel grid, and a pixel-wise adaptive layer normalization (adaLN)~\cite{Ma_2026_CVPR_DeCo, Yu_2026_CVPR_PixelDiT} predicts modulation parameters for every pixel.
Transplanting this per-pixel modulation to video is infeasible at our scale: the activation memory scales with the full-resolution $T\times H\times W$ volume, and our target setting of 32 frames at 1080p already exhausts an 80\,GB GPU.
We therefore run the local branch on a grid downsampled $8\times$ spatially and $2\times$ temporally, still finer than the DiT token grid but coarse enough to stay within memory.
The DiT output is upsampled to this grid as $\tilde{s}=\mathrm{Up}(s)$, and the \emph{Patch Modulation} block predicts one set of adaLN parameters per grid cell rather than a distinct set for every output pixel,
\begin{equation}
    (\gamma, \beta, \alpha) = \mathrm{MLP}_{\text{mod}}(\tilde{s}),
    \label{eq:arch-patchmod-params}
\end{equation}
\begin{equation}
    h \leftarrow h + \alpha \odot \mathrm{MLP}_{\text{feat}}\!\big( (1+\gamma)\odot \mathrm{LN}(h) + \beta \big),
    \label{eq:arch-patchmod-update}
\end{equation}
where $h$ is the local feature stream on this grid, initialized by patch-embedding the noisy $X_t$ and matched in resolution to $\tilde{s}$, and $\mathrm{LN}$ denotes layer normalization.
This keeps memory tractable while injecting the upsampled DiT context $\tilde{s}$ into $h$, and improves markedly over the pure-DiT baseline (Fig.~\ref{fig:3_3}).

Patch Modulation injects the global feature $\tilde{s}$ independently at each local-grid cell, but provides limited explicit interaction among neighboring spatial and temporal locations.
At this finer grid, convolution provides linear-complexity local mixing and an appropriate inductive bias for recovering spatial texture and short-range temporal variation.
We therefore interleave Patch Modulation with a lightweight \emph{Spatial--Temporal Conv} branch that factorizes a 3D convolution into a 2D spatial convolution $\mathrm{SConv}$ followed by a 1D temporal convolution $\mathrm{TConv}$:
\begin{equation}
    h \leftarrow h + \mathrm{TConv}\!\big( \mathrm{SConv}(h) \big),
    \label{eq:arch-stconv}
\end{equation}
Each \emph{Patch Refine} block combines Patch Modulation with a Spatial--Temporal Conv module containing three factorized convolutional layers.
Finally, following the DCVC line~\cite{jia2025towards,li2026ultra}, the \emph{Frame Refine} head restores the temporal and spatial resolution of $h$ and maps it to the RGB reconstruction $\hat{X}_0$ (Fig.~\ref{fig:3_2}).
The detailed ablation of each component is provided in Sec.~\ref{sec:exp-ablation}, which demonstrates the effectiveness of the individual design choices.

\begin{figure*}[t]
    \centering
    \includegraphics[width=1.0\linewidth]{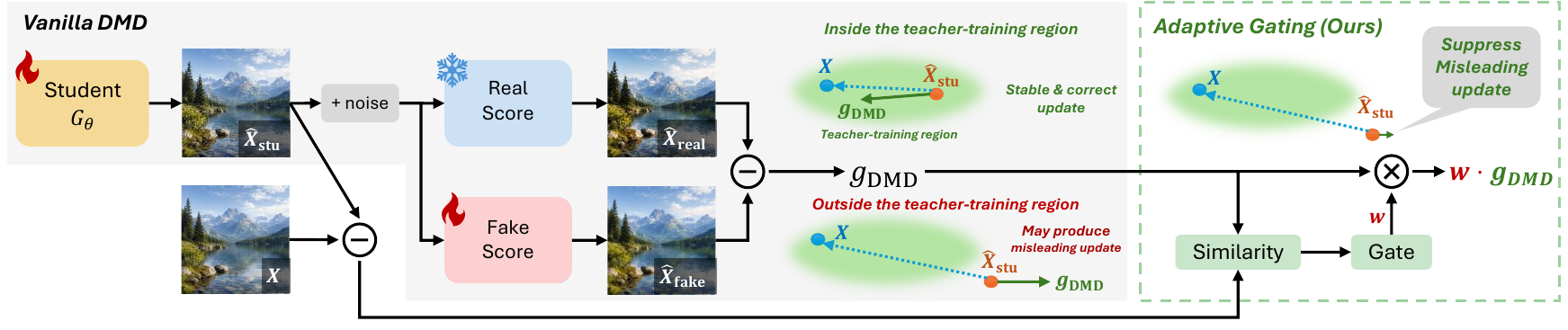}
    \vspace{-7mm}
    \caption{
        \textbf{Teacher-side guidance failure and Adaptive Score Distillation.}
        Left: vanilla DMD produces the update $g_{\text{DMD}}$.
        Middle: the green area denotes the region covered during teacher training. When student reconstructions remain close to this region, teacher guidance yields stable, corrective DMD updates; substantial deviations may instead yield misleading updates.
        Right: our alignment-based gate attenuates such misleading updates.
    }
    \label{fig:4_1}
    \vspace{-4mm}
\end{figure*}

\subsection{Training Objective}
\label{sec:pixvdcm-train}

We initialize the diffusion model from scratch and train it jointly with the contextual codec.
Consequently, the compression objective shapes both the decoded condition $c$ and the conditional reconstruction model.
The overall objective combines the flow-matching term in Eq.~\eqref{eq:fm-loss} with an auxiliary-decoder distortion term and a rate term from the codec part:
\begin{equation}
    \mathcal{L} = L_{\mathrm{fm}} + \lambda_{\text{aux}}\,\mathcal{L}_{\text{aux}} + \lambda_{R}\,\mathcal{L}_{R}.
    \label{eq:pixvdcm-train-loss}
\end{equation}
Here $L_{\mathrm{fm}}$ supervises the diffusion transport in pixel space, $\mathcal{L}_{\text{aux}}$ is an auxiliary-decoder distortion term, and $\mathcal{L}_{R}$ is the estimated coding rate under the learned entropy model.

Following the compression-oriented design in~\cite{Jia_2026_CVPR}, where this auxiliary objective is shown to be effective, $\mathcal{L}_{\text{aux}}$ directly supervises the reconstruction that the \emph{Aux Decoder} produces from the condition $c$ (Fig.~\ref{fig:3_1}) with MSE and LPIPS against the ground-truth GOP.
This gives the codec a direct, perceptually grounded signal for retaining meaningful structure in $c$, rather than relying solely on gradients back-propagated through the diffusion branch.
The resulting multi-step model is a strong generative video codec in its own right; its diffusion model is subsequently frozen as the teacher for one-step distillation.


\section{One-Step Diffusion with Adaptive Score Distillation}
\label{sec:pixvdc1}

The compression-oriented model developed in Sec.~\ref{sec:pixvdcm} is a complete generative video codec, but its multi-step ODE sampler limits decoding throughput.
We therefore use its diffusion model as the frozen teacher for the one-step GenVC, following the growing use of one-step diffusion in image and video codecs~\cite{NEURIPS2025_352a67c0,Jia_2026_CVPR,jia2026codliterealtimediffusionbasedgenerative,ma2025diffvc,ma2026diffvc,xue2026single}.
Although some existing one-step diffusion codecs~\cite{NEURIPS2025_352a67c0,Jia_2026_CVPR,jia2026codliterealtimediffusionbasedgenerative} directly adopt DMD, applying it in our setting drives the student toward motion-stalled reconstructions.
We trace this behavior to a teacher-side guidance failure and introduce Adaptive Score Distillation to address it.
The remainder of this section reviews DMD (Sec.~\ref{sec:pixvdc1-prelim}), analyzes this failure (Sec.~\ref{sec:pixvdc1-failure}), and presents our distillation method (Sec.~\ref{sec:pixvdc1-asd}).

\subsection{DMD Preliminaries}
\label{sec:pixvdc1-prelim}
We review the distribution-matching component of DMD that our method builds upon.
For each source input $X$, let $c_{\text{S}}$ and $c_{\text{T}}$ denote the input conditions of the student and teacher diffusion models, respectively.
The one-step student maps a Gaussian noise sample $\epsilon$ and $c_{\text{S}}$ to a reconstruction $\hat{X}_{\text{stu}} = G_\theta(\epsilon, c_{\text{S}})$.
DMD matches the perturbed student distribution $p_{\text{stu},t}(\cdot\mid c_{\text{S}})$ to the perturbed target distribution $p_{\text{real},t}(\cdot\mid c_{\text{T}})$ by minimizing their reverse Kullback--Leibler divergence over noise levels $t$.
For a student reconstruction, we sample independent perturbation noise $\epsilon'$ and form $X_t=(1-t)\hat{X}_{\text{stu}}+t\epsilon'$, following the data-to-noise convention of Sec.~\ref{sec:pixvdcm-bg}.
Conditioned on $c_{\text{T}}$, the perturbed sample is evaluated by two score networks: the frozen multi-step teacher serves as the \emph{real} score network, while an online \emph{fake} score network tracks the current student distribution (Fig.~\ref{fig:4_1}, left).
Writing their score estimates as $s_{\text{real}}(X_t,t\mid c_{\text{T}})$ and $s_{\text{fake}}(X_t,t\mid c_{\text{T}})$, the DMD gradient takes the form, up to a positive noise-level weight,
\begin{equation}
    \nabla_\theta \mathcal{L}_{\text{DMD}}
    \;\propto\;
    \mathbb{E}\!\left[\big(s_{\text{fake}}-s_{\text{real}}\big)\,\partial_\theta\hat{X}_{\text{stu}}\right].
    \label{eq:pixvdc1-dmd-kl}
\end{equation}
Both networks are parameterized to predict clean samples, denoted $\hat{X}_{\text{real}}$ and $\hat{X}_{\text{fake}}$.
Because $s_{\text{fake}}-s_{\text{real}}$ is proportional to $\hat{X}_{\text{fake}}-\hat{X}_{\text{real}}$, gradient descent moves the student along the update direction
\begin{equation}
    g_{\text{DMD}} = \hat{X}_{\text{real}}-\hat{X}_{\text{fake}}.
    \label{eq:pixvdc1-gdmd}
\end{equation}
We train the fake score network by flow matching on student outputs, alternating its updates with those of the student.

\begin{figure}[t]
    \centering
    \includegraphics[width=1.0\linewidth]{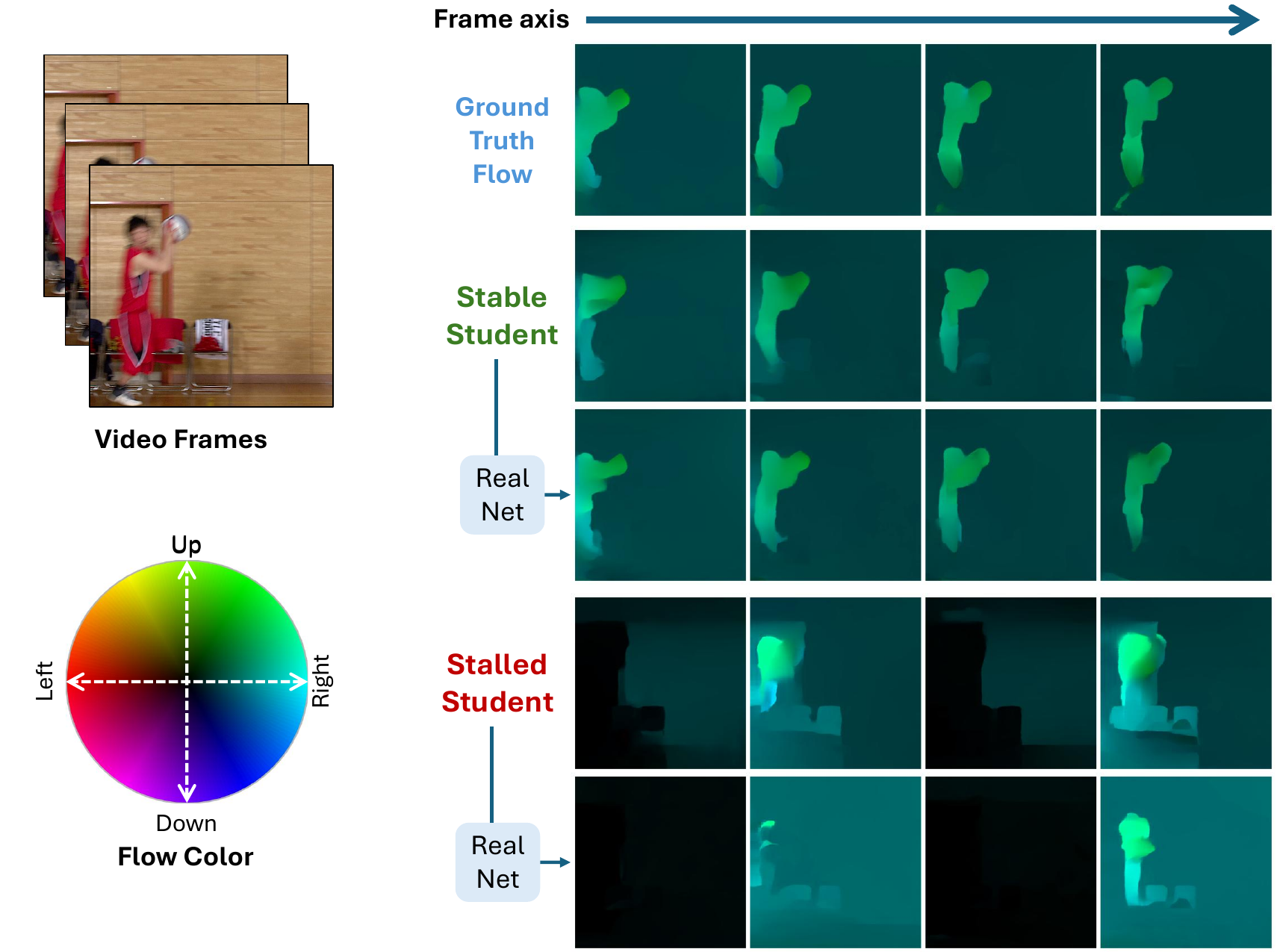}
    \vspace{-7mm}
    \caption{
        \textbf{Visual evidence of the teacher-side guidance failure.}
        Optical flow of the ground truth and the stable/stalled student reconstructions.
        Perturbed versions of both student outputs are passed through the same frozen teacher: motion from the stable student is largely preserved, whereas the already degraded motion of the stalled student is suppressed further.
    }
    \label{fig:4_2}
    \vspace{-4mm}
\end{figure}

\subsection{Teacher-Side Guidance Failure}
\label{sec:pixvdc1-failure}

Applying the DMD recipe of Sec.~\ref{sec:pixvdc1-prelim} directly to our multi-step teacher does not transfer cleanly to the one-step video setting.
As distillation proceeds, the reconstruction quality degrades, most visibly in motion: the student produces motion-stalled reconstructions.
Fig.~\ref{fig:4_2} visualizes the optical flow between consecutive frames, color-coded by direction with brighter regions marking larger motion.
The ground-truth flow forms a bright, coherent region that follows the moving player, whereas the stalled student's flow is dim and fragmented, leaving the player nearly frozen.

\begin{figure}[t]
    \centering
    \includegraphics[width=1.0\linewidth]{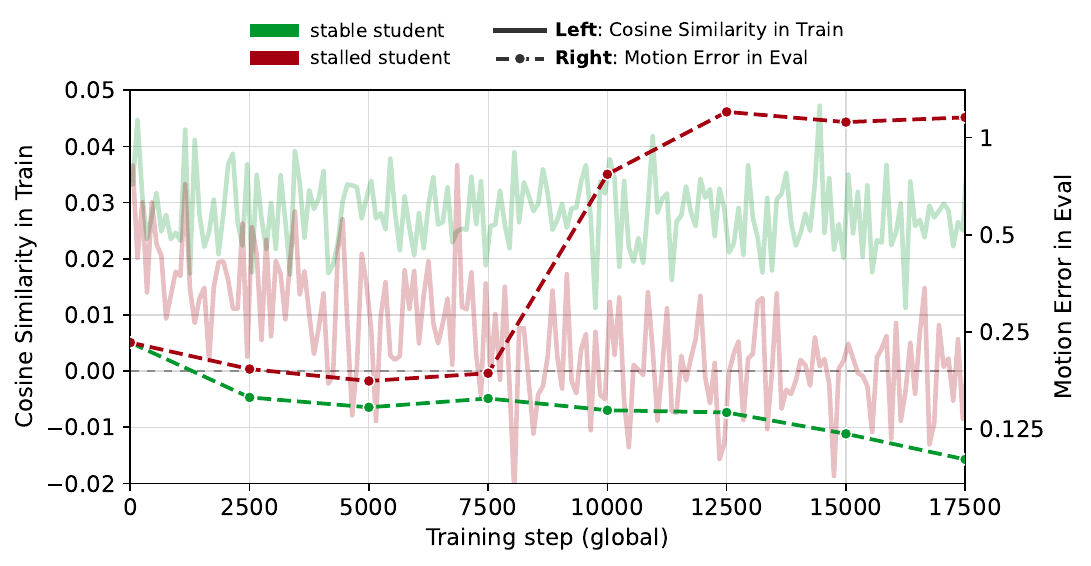}
    \vspace{-7mm}
    \caption{
        \textbf{Training dynamics of the teacher-side guidance failure.}
        Training-time cosine (solid, left axis) and evaluation-time Motion Error (dashed, right axis) for the stalled student (red) and the stable student (green).
    }
    \label{fig:4_3}
    \vspace{-4mm}
\end{figure}

To track this failure quantitatively, we measure a \emph{motion error} (ME), the endpoint error between the optical flow of the reconstruction and that of the ground truth over consecutive frame pairs, normalized by the ground-truth flow magnitude,\footnote{The same absolute flow error is large on a near-static clip yet negligible on a fast-moving one; dividing by the ground-truth flow magnitude makes ME a dimensionless relative error that is comparable across clips.}
\begin{equation}
    \mathrm{ME} = \frac{\sum_{i=1}^{T-1} \big\| \mathcal{F}(\hat{X}_i, \hat{X}_{i+1}) - \mathcal{F}(X_i, X_{i+1}) \big\|}{\sum_{i=1}^{T-1} \big\| \mathcal{F}(X_i, X_{i+1}) \big\|},
    \label{eq:pixvdc1-me}
\end{equation}
where $\mathcal{F}$ is the pre-trained RAFT optical-flow model~\cite{teed2020raft}, $T$ is the number of frames; $X_i$ and $\hat{X}_i$ are the $i$-th ground-truth and reconstructed frames.
Tracked across distillation (Fig.~\ref{fig:4_3}, stalled student), the motion error confirms that the collapse is a training-dynamics effect rather than a poor initialization: it stays low through an extended early phase, then surges sharply to a high plateau.

We trace this failure to the frozen teacher's response to student-induced inputs.
The resulting perturbations can move beyond the regions covered by the ground-truth perturbations used to train the teacher.
On such inputs, the teacher can produce a plausible but even lower-motion $\hat{X}_{\text{real}}$ instead of restoring the missing motion.
Because the fake score network tracks the current student, the resulting update $g_{\text{DMD}}$ can then point away from the ground-truth direction and reinforce the drift that produced it (Fig.~\ref{fig:4_1}).
We refer to this input-dependent loss of corrective guidance as a \emph{teacher-side guidance failure}.

We isolate this input dependence by passing perturbed reconstructions from a stable student and a stalled student through the same teacher (Fig.~\ref{fig:4_2}).
With identical weights, the teacher largely preserves motion for stable-student inputs but further suppresses motion for stalled-student inputs, showing that the failure arises from shifted student inputs rather than an inherent tendency of the teacher to suppress motion.

Since paired source videos are available during codec training, we can assess online whether each DMD update remains source-corrective.
For each sample $b$, we measure the cosine alignment between the DMD update and the direction from the current reconstruction toward the ground truth,
\begin{equation}
    \cos_{\text{gt}}^{(b)}
    = \cos\!\big(g_{\text{DMD}}^{(b)},\; X^{(b)}-\hat{X}_{\text{stu}}^{(b)}\big),
    \label{eq:pixvdc1-cosgt}
\end{equation}
where $b$ indexes samples in the batch and $X$ is the ground-truth video.
A positive value indicates that the update has a component toward the source, while a negative value indicates that it actively pushes the reconstruction away.
For the stalled student, the batch-averaged cosine falls from a small positive value and crosses zero before ME rises sharply; for the stable student, it remains positive and ME stays low (Fig.~\ref{fig:4_3}).
The cosine therefore provides an online early indicator of impending stalling without requiring optical-flow evaluation.
This per-sample, ground-truth-referenced signal is exactly what our method gates on, which we develop next.

\subsection{Adaptive Score Distillation}
\label{sec:pixvdc1-asd}

Section~\ref{sec:pixvdc1-failure} established $\cos_{\text{gt}}^{(b)}$ as a per-sample, online indicator of whether a DMD update still points toward the ground truth.
To break the feedback loop, we turn this signal into a per-sample gate: each sample $b$ in a batch has its update weighted by a smooth function of its alignment,
\begin{equation}
    w_b = \sigma\!\left( \frac{\cos_{\text{gt}}^{(b)} - \tau}{\kappa} \right),
    \label{eq:pixvdc1-gate}
\end{equation}
with $\sigma(\cdot)$ the logistic sigmoid, $\tau$ a threshold, and $\kappa$ a temperature.
Better-aligned updates receive larger weight and drive distribution matching, whereas opposing ones are suppressed toward zero.
The sigmoid attenuates borderline updates gradually instead of switching them off at a hard threshold.
Treating $w_b$ as a stop-gradient coefficient, this reweighting enters the DMD gradient per sample,
\begin{equation}
    \nabla_\theta \mathcal{L}_{\text{DMD}}
    \;\propto\;
    -\mathbb{E}\!\left[w_b^2\,g_{\text{DMD}}^{(b)}\,\partial_\theta\hat{X}_{\text{stu}}^{(b)}\right],
    \label{eq:pixvdc1-gated}
\end{equation}
so gradient descent moves each student sample along $w_b^2 g_{\text{DMD}}^{(b)}$.
Poorly aligned samples contribute little, while well-aligned samples continue to drive distribution matching.
Since $w_b^2$ remains monotonic in the alignment, the square preserves the gating behavior while sharpening its attenuation.
This per-sample reweighting is the only change to the standard DMD student gradient and constitutes our \emph{Adaptive Score Distillation} (Fig.~\ref{fig:4_1}, right).
We inject this gated direction through the stop-gradient surrogate
\begin{equation}
    \mathcal{L}_{\text{DMD}}
    = \frac{1}{2BP}\sum_{b=1}^{B} w_b^2
    \left\|\hat{X}_{\text{stu}}^{(b)}-
    \mathrm{sg}\!\left(\hat{X}_{\text{stu}}^{(b)}+g_{\text{DMD}}^{(b)}\right)\right\|_2^2,
    \label{eq:pixvdc1-gated-loss}
\end{equation}
where $B$ is the batch size and $P$ is the number of elements in each sample.
This yields $\nabla_{\hat{X}_{\text{stu}}^{(b)}}\mathcal{L}_{\text{DMD}}
\propto -w_b^2 g_{\text{DMD}}^{(b)}$.

Following prior one-step codecs~\cite{NEURIPS2025_352a67c0,Jia_2026_CVPR}, we first initialize the student from the multi-step teacher and jointly optimize it with the codec under $\mathcal{L}_{\text{init}} = \mathcal{L}_{\text{dist}} + \lambda_{R}\,\mathcal{L}_{R}$, where $\mathcal{L}_{R}$ is the rate term and $\mathcal{L}_{\text{dist}}$ is defined below.
We then freeze the codec and fine-tune the diffusion student while updating the fake score network online; the objective below therefore contains no rate term.
The gated DMD term enters the one-step distillation objective together with distortion losses,
\begin{equation}
    \mathcal{L} = \lambda_{\text{D}}\, \mathcal{L}_{\text{DMD}} + \mathcal{L}_{\text{dist}},
    \label{eq:pixvdc1-distill}
\end{equation}
where $\mathcal{L}_{\text{DMD}}$ is defined in Eq.~\eqref{eq:pixvdc1-gated-loss}.
The source-referenced distortion term is
\begin{equation}
    \mathcal{L}_{\text{dist}} = \mathcal{L}_{\text{pix}} + \lambda_{\text{L}}\,\mathcal{L}_{\text{LPIPS}} + \lambda_{\text{F}}\,\mathcal{L}_{\text{flow}},
    \label{eq:pixvdc1-dist}
\end{equation}
where $\mathcal{L}_{\text{pix}}$ is the mean $\ell_1$ reconstruction error, $\mathcal{L}_{\text{LPIPS}}$ measures perceptual fidelity, and $\mathcal{L}_{\text{flow}}$ is the mean $\ell_1$ discrepancy between frozen-RAFT flow fields of corresponding source and reconstructed frame pairs.
The two components are complementary: adaptive gating suppresses poorly aligned DMD updates while preserving reliable distribution-matching guidance that improves one-step reconstruction realism, whereas $\mathcal{L}_{\text{dist}}$ anchors the reconstruction to the source video.
Algorithm~\ref{alg:asd} further illustrates the distillation training step.

\begin{algorithm}[t]
\caption{Adaptive Score Distillation: one training step}
\label{alg:asd}
\small
\begin{algorithmic}[1]
\Require student $G_\theta$; frozen teacher $D_{\text{T}}$; fake score network $D_\phi$; training tuples $(c_{\text{S}},c_{\text{T}},X)$; gate $\tau,\kappa$; weight $\lambda_{\text{D}}$; fake-score update count $n_{\text{fake}}$
\For{$i = 1$ \textbf{to} $n_{\text{fake}}$} \Comment{update the fake score network}
    \State draw a fresh batch $(c_{\text{S}},c_{\text{T}},X)$;\; $\epsilon \sim \mathcal{N}(\mathbf{0}, \mathbf{I})$
    \State $\hat{X}_{\text{stu}} \gets G_\theta(\epsilon, c_{\text{S}})$ \Comment{one-step reconstruction}
    \State sample $t_\phi$ and $\xi_\phi\sim\mathcal{N}(\mathbf{0},\mathbf{I})$
    \State $X_{t_\phi}\gets(1-t_\phi)\mathrm{sg}(\hat{X}_{\text{stu}})+t_\phi\xi_\phi$
    \State $\hat{X}_{\text{fake}}\gets D_\phi(X_{t_\phi},t_\phi,c_{\text{T}})$
    \State $v_{\text{tgt}}\gets\xi_\phi-\mathrm{sg}(\hat{X}_{\text{stu}})$
    \State $v_\phi\gets(X_{t_\phi}-\hat{X}_{\text{fake}})/t_\phi$
    \State update $\phi$ using $\|v_\phi-v_{\text{tgt}}\|_2^2$
\EndFor
\State draw a fresh batch $(c_{\text{S}},c_{\text{T}},X)$;\; $\epsilon \sim \mathcal{N}(\mathbf{0}, \mathbf{I})$ \Comment{student update}
\State $\hat{X}_{\text{stu}} \gets G_\theta(\epsilon, c_{\text{S}})$
\State sample $t$ and $\xi\sim\mathcal{N}(\mathbf{0},\mathbf{I})$
\State $X_t\gets(1-t)\hat{X}_{\text{stu}}+t\xi$
\State $\hat{X}_{\text{real}} \gets D_{\text{T}}^{\text{CFG}}(X_t,t,c_{\text{T}})$; \: $\hat{X}_{\text{fake}} \gets D_\phi(X_t,t,c_{\text{T}})$
\For{each sample $b$} \Comment{\textbf{ours}}
    \State $d_b \gets \operatorname{mean}|\hat{X}_{\text{stu}}^{(b)}-\hat{X}_{\text{real}}^{(b)}|$ \Comment{per-sample norm}
    \State $g_{\text{DMD}}^{(b)} \gets (\hat{X}_{\text{real}}^{(b)}-\hat{X}_{\text{fake}}^{(b)})/(d_b+10^{-4})$
    \State $\cos_{\text{gt}}^{(b)} \gets \cos\!\big(g_{\text{DMD}}^{(b)},\, X^{(b)}-\hat{X}_{\text{stu}}^{(b)}\big)$
    \State $w_b \gets \mathrm{sg}\!\left[\sigma\!\big((\cos_{\text{gt}}^{(b)} - \tau)/\kappa\big)\right]$
\EndFor
\State $\mathcal{L}_{\text{DMD}} \gets \dfrac{1}{2BP}\sum_b w_b^2\left\|\hat{X}_{\text{stu}}^{(b)}-\mathrm{sg}\!\big(\hat{X}_{\text{stu}}^{(b)}+g_{\text{DMD}}^{(b)}\big)\right\|_2^2$
\State compute $\mathcal{L}_{\text{dist}}(\hat{X}_{\text{stu}},X)$ \Comment{Eq.~\eqref{eq:pixvdc1-dist}}
\State $\mathcal{L} \gets \lambda_{\text{D}}\,\mathcal{L}_{\text{DMD}} + \mathcal{L}_{\text{dist}}$ \Comment{Eq.~\eqref{eq:pixvdc1-distill}}
\State update $\theta$ by $\nabla_\theta \mathcal{L}$
\end{algorithmic}
\end{algorithm}

Adaptive gating breaks the observed feedback loop and restores stable distillation in our setting.
With the gate in place, the training-time $\cos_{\text{gt}}$ stays positive and the evaluation-time motion error stays low throughout, rather than the zero-crossing and motion-error surge seen without it (Fig.~\ref{fig:4_3}, green vs red).
The one-step student thereby avoids the motion-stalling regime observed with direct DMD.


\begin{figure*}[t]
    \centering
    \includegraphics[width=1.0\linewidth]{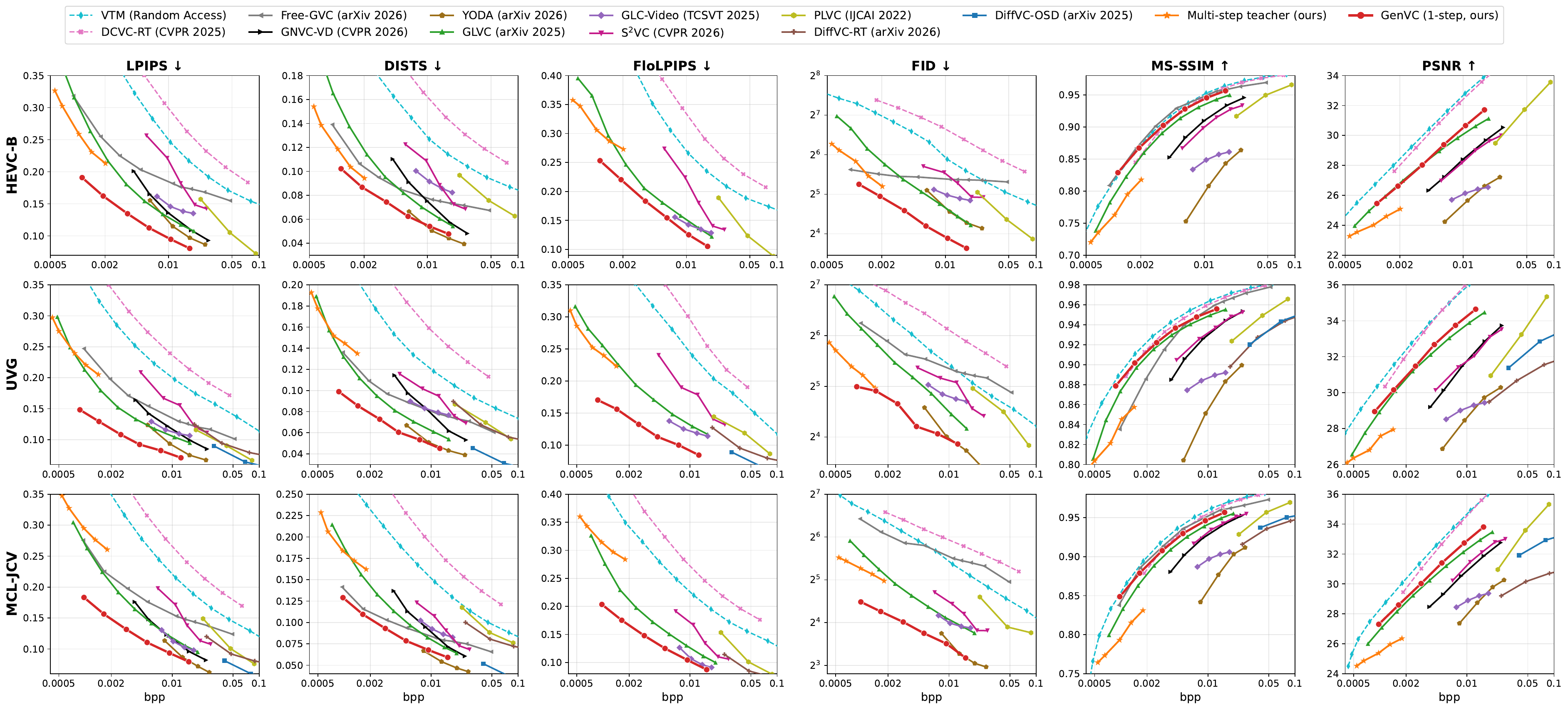}
    \vspace{-7mm}
    \caption{
        Rate--quality performance on HEVC Class~B, UVG, and MCL-JCV.
        The multi-step model is evaluated as a standalone codec, while GenVC denotes the distilled one-step system.
        Zoom in for a better view.
    }
    \label{fig:5_1}
    \vspace{-4mm}
\end{figure*}

\section{Experiments}
\label{sec:exp}

\subsection{Experimental Setup}
\label{sec:exp-setup}

\textbf{Implementation Details.}
We adopt the contextual codec backbone of DCVC-UF~\cite{li2026ultra} and initialize only this codec component from its pretrained checkpoint; the multi-step diffusion model is initialized from scratch and jointly optimized with the codec.
We set $K=4$, such that each GOP comprises four 8-frame chunks (32 frames).
We train GenVC in three stages.
In the first stage, we train the multi-step teachers for total 12 epochs under the objective of Eq.~\eqref{eq:pixvdcm-train-loss}, using $\lambda_{R} \in \{0.56, 0.80, 1.12, 1.76, 2.40\}$ to obtain five models at different bitrates.
In the second stage, we jointly train each one-step student and its codec under the initialization objective $\mathcal{L}_{\text{init}}$ introduced in Sec.~\ref{sec:pixvdc1}. We sweep $\lambda_{R} \in \{10.24, 5.12, 2.56, 1.28, 0.64, 0.32\}$ to set six bitrates and train each model for 3 epochs; five models reuse the corresponding first-stage teachers, while the sixth uses the highest-bitrate teacher.
In the third stage, we freeze the student codec and fine-tune the one-step student for 1 epoch with Adaptive Score Distillation (Eq.~\eqref{eq:pixvdc1-distill}), while updating the auxiliary fake score network online; the corresponding teacher codec remains frozen and supplies $c_{\text{T}}$.
We set $\lambda_{\text{aux}}=0.01$, $\lambda_{\text{L}}=1.0$, $\lambda_{\text{F}}=2.0$, and $\lambda_{\text{D}}=8.0$ in training objectives.
We use gate threshold $\tau=0.015$ and temperature $\kappa=0.02$, and perform $n_{\text{fake}}=4$ fake-score updates on fresh batches per student update.
Writing $\rho=1-t$ for the data coefficient under the convention of Sec.~\ref{sec:pixvdcm-bg}, we draw $\rho$ by applying a sigmoid to a standard Gaussian and rescale it to $[0.05,0.95]$ for the student-side DMD update; multi-step teacher training and fake-score training use the unrescaled logit-normal distribution over $\rho\in(0,1)$.
The frozen teacher prediction uses classifier-free guidance (CFG) at scale 1.5, whereas the fake prediction is unguided.
At inference, the multi-step codec uses a 20-step Euler sampler with the same CFG scale, whereas GenVC uses one-step inference.
Our models are trained on four A100 80GB GPUs with the AdamW optimizer~\cite{loshchilov2017decoupled}.

\textbf{Datasets.}
We train our models on the OpenVid dataset~\cite{nan2024openvid}, with clips randomly cropped to a spatial resolution of $512 \times 512$ and spanning three consecutive GOPs (96 frames in total).
For evaluation, we use the UVG~\cite{mercat2020uvg}, MCL-JCV~\cite{wang2016mcl}, and HEVC Class~B~\cite{sullivan2012overview} datasets, testing on the first 96 frames of each sequence at $1920 \times 1080$.
All YUV frames are converted to RGB following the ITU-R BT.709 standard.

\textbf{Compared Methods.}
We compare GenVC against three groups of baselines:
(i) the traditional codec VTM under the random-access configuration with a 32-frame GOP~\cite{bross2021overview};
(ii) the distortion-oriented neural video codec DCVC-RT~\cite{jia2025towards}; and
(iii) generative video codecs: PLVC~\cite{yang2022perceptual}, S$^2$VC~\cite{xue2026single}, GLC-video~\cite{qi2025generative}, GLVC~\cite{guo2025generative}, YODA~\cite{li2026yoda}, GNVC-VD~\cite{mao2026generative}, Free-GVC~\cite{ling2026free}, DiffVC-OSD~\cite{ma2025diffvc}, and DiffVC-RT~\cite{ma2026diffvc}.

\textbf{Metrics.} 
We evaluate quality primarily with perceptual metrics:
LPIPS~\cite{zhang2018unreasonable} and DISTS~\cite{ding2020image} for fidelity, FID~\cite{heusel2017gans} for distributional realism, and FloLPIPS~\cite{danier2022flolpips} for motion-aware quality.
We additionally report PSNR and MS-SSIM~\cite{wang2003multiscale} as objective distortion references.
Rate is measured in bits per pixel (bpp), and BD-Rate~\cite{bjontegaard2001calculation} summarizes bitrate savings at matched perceptual quality.
For ablations, the analogous BD-Distortion reports the average metric difference at matched rate over the shared rate range.
For complexity, we report throughput in frames per second (fps) at $1920 \times 1080$.
The motion error (ME) defined in Sec.~\ref{sec:pixvdc1-failure} is reused as a diagnostic in the ablation study.

\subsection{Main Results}
\label{sec:exp-main}

\textbf{Standalone Multi-Step Codec.}
Fig.~\ref{fig:5_1} reports the rate--quality curves of both the multi-step model and GenVC across all six metrics.
Before distillation, the multi-step model already provides competitive perceptual reconstruction at ultra-low rates on all three datasets.
This result establishes that a video diffusion model trained from scratch jointly with the codec can serve as a strong generative codec in its own right, rather than only as an intermediate teacher.
Its remaining limitation is iterative sampling, whose cost is quantified below.

\textbf{Distilled One-Step Codec.}
After one-step distillation, GenVC consistently lies on or near the leading perceptual rate--quality frontier, with its clearest gains on LPIPS and FID.
Relative to GLVC, GenVC saves 59.6\%, 69.0\%, and 59.0\% bitrate at matched LPIPS on HEVC Class~B, UVG, and MCL-JCV, respectively, averaging 62.5\%.
At matched FID, the corresponding savings are 69.6\%, 69.5\%, and 74.9\%, averaging 71.3\%.
The decoded frames in Fig.~\ref{fig:2} qualitatively corroborate these gains: GenVC reconstructs fine texture and detail at the lowest bitrate among the compared examples.
GenVC also achieves competitive rate--quality performance on FloLPIPS and DISTS across the three datasets.
As expected from the rate--perception trade-off, it trails the distortion-oriented VTM and DCVC-RT on PSNR and MS-SSIM.
Nevertheless, GenVC remains competitive with other generative codecs on both metrics, demonstrating strong objective fidelity alongside its perceptual gains.
Overall, GenVC's perceptual rate--quality gains over codecs adapted from large pretrained generative backbones support the effectiveness of learning the diffusion model specifically for compression.

\textbf{Relation to the Multi-Step Teacher.}
Across most of their shared rate range, GenVC further improves upon the multi-step model.
The student is not trained to directly reproduce the teacher's sampled reconstructions.
Instead, DMD uses score estimates from the frozen teacher to guide the student toward distributionally realistic outputs, while the source-referenced losses optimize fidelity to the paired video.
Adaptive Score Distillation additionally attenuates non-corrective teacher guidance.
The teacher's sampled rate--quality curve is therefore a reference rather than a strict upper bound on the separately optimized one-step codec.

\textbf{Throughput and Complexity.}
Our multi-step teacher uses 20-step CFG sampling and reaches 0.40~fps at 1080p on an A100, as reported in Table~\ref{tab:complexity}.
On the same device, one-step distillation raises decoding to 15.1~fps with the same 578.4M total parameter count and unchanged encoding throughput, a $37.8\times$ speedup over the teacher.
Like prior DMD-distilled one-step codecs~\cite{NEURIPS2025_352a67c0,jia2026codliterealtimediffusionbasedgenerative}, GenVC requires no CFG at inference; the measured speedup therefore reflects both the reduction to one-step inference and the removal of CFG.
Despite differences in evaluation hardware, GenVC achieves the highest decoding throughput among the compared diffusion-based codecs, surpassing YODA, S$^2$VC, GNVC-VD, and the DiffVC series.
Together with the BD-Rate results, these findings show that compression-oriented diffusion can deliver SOTA perceptual quality and efficient decoding at a sub-billion scale, using substantially fewer parameters than codecs built on billion-scale pretrained backbones.

\begin{table}[t]
\centering
\scriptsize
\setlength{\tabcolsep}{1pt}
\caption{Parameter count, throughput, and LPIPS BD-Rate on MCL-JCV.}
\vspace{-2mm}
\label{tab:complexity}
\begin{tabular*}{\columnwidth}{@{\extracolsep{\fill}}llcccc@{}}
\toprule
Device & Method & Params (M) & Enc.\ (fps) & Dec.\ (fps) & BD-Rate$\downarrow$ \\
\midrule
\multirow{3}{*}{A800 (80GB)} & DiffVC-OSD~\cite{ma2025diffvc} & 1488.7 & 0.7   & 0.9  & -- \\
                             & DiffVC-RT~\cite{ma2026diffvc}  & 474.3  & 105.9 & 7.7  & 142.6\% \\
                             & GNVC-VD~\cite{mao2026generative} & 2334.5 & 6.5 & 0.6 & -1.5\% \\
\midrule
RTX 5090 & YODA~\cite{li2026yoda} & 1063.2 & --    & 1.0  & -37.1\% \\
\midrule
\multirow{4}{*}{A100 (80GB)} & GLVC~\cite{guo2025generative} & 227.3 & 5.2   & 3.3  & 0.0\% \\ 
                             & S$^2$VC~\cite{xue2026single} & 1346.0 & 6.6   & 1.3  & 134.6\% \\
                             & Teacher (ours) & 578.4  & 309.1  & 0.4 & -- \\
                             & \textbf{GenVC (ours)} & 578.4  & 309.1 & \textbf{15.1} & \textbf{-59.0\%} \\
\bottomrule
\end{tabular*}\par
\vspace{1mm}
\noindent\parbox[t]{\columnwidth}{\scriptsize
GenVC and its teacher: 100.4M (codec) + 478.0M (diffusion) = 578.4M\par
Resolution for fps calculation: 1920$\times$1080; BD-Rate anchor: GLVC.\par
\textcolor{gray}{For coding speed, ``--'' denotes no data; for BD-Rate, ``--'' denotes no overlap.}\par
\textcolor{gray}{DiffVC-OSD, DiffVC-RT, and YODA fps are from~\cite{ma2026diffvc}; GNVC-VD fps are from~\cite{mao2026generative}; A100 fps are measured by us.}\par
}
\par
\vspace{-5mm}
\end{table}

\begin{figure}[t]
    \centering
    \includegraphics[width=1.0\linewidth]{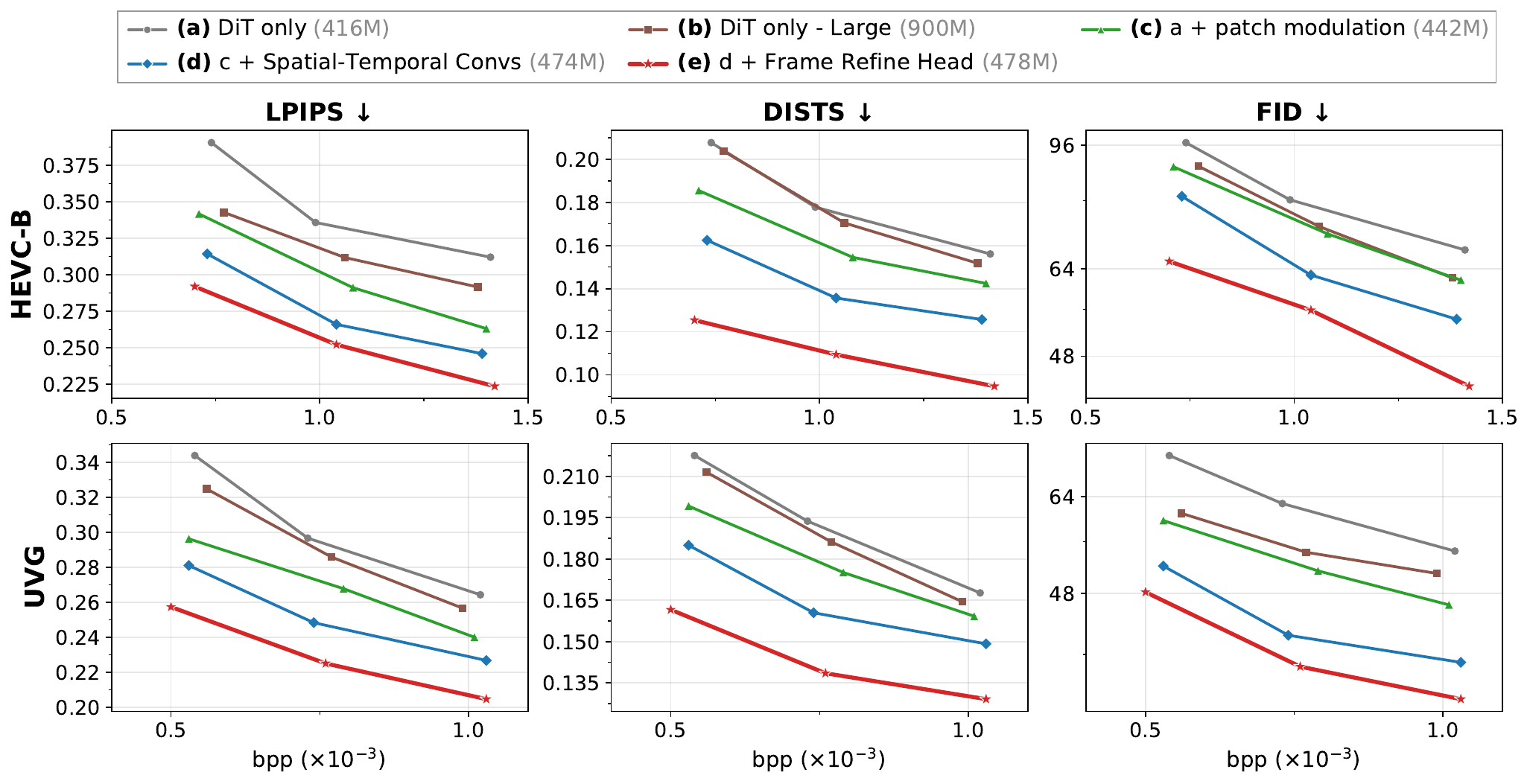}
    \vspace{-8mm}
    \caption{
        Architecture ablation of the multi-step teacher, with the parameter count of each diffusion variant annotated in the legend.
        The standard per-pixel modulation is omitted because it exceeds an 80\,GB GPU memory budget under the 32-frame 1080p configuration.
    }
    \label{fig:5_2}
    \vspace{-6mm}
\end{figure}

\subsection{Ablation Study}
\label{sec:exp-ablation}

\textbf{Global-to-Local Architecture.}
We first evaluate the progressive architecture decisions in Sec.~\ref{sec:pixvdcm-arch}.
As shown by the full rate--quality curves in Fig.~\ref{fig:5_2}, adding patch-level modulation, Spatial--Temporal Convs, and the Frame Refine head successively improves LPIPS, DISTS, and FID on both datasets.
Fig.~\ref{fig:3_3} compares architectural refinement with model scaling: on HEVC Class~B, enlarging the DiT-only baseline from 416M to 900M parameters reduces BD-LPIPS and BD-FID by only 0.024 and 4.75, whereas the complete 478M hierarchy reduces them by 0.087 and 25.51.
These results indicate that the gains arise primarily from the global-to-local refinement hierarchy rather than parameter count alone, validating the effectiveness of our video-specific design.

\begin{figure}[t]
    \centering
    \includegraphics[width=1.0\linewidth]{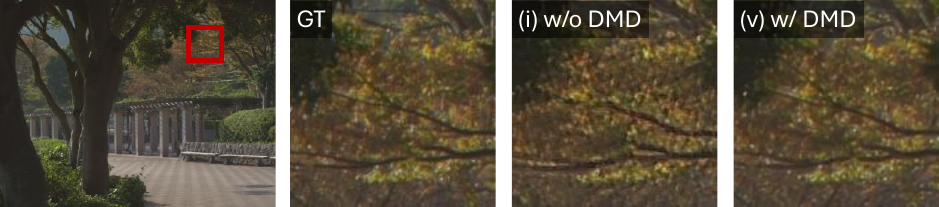}
    \vspace{-7mm}
    \caption{
        \textbf{Qualitative contribution of DMD.}
        Compared with the distortion-only variant~(i), the full model~(v) with gated DMD suppresses structured artifacts and reconstructs finer texture closer to the ground truth.
    }
    \label{fig:dmd-visual}
    \vspace{-5mm}
\end{figure}

\textbf{Distribution Matching Distillation.}
Our central ablation dissects the roles of the DMD term, the adaptive gate, and the distortion loss in the one-step distillation stage, with $\mathcal{L}_{\text{dist}}$ defined in Eq.~\eqref{eq:pixvdc1-dist}.
Table~\ref{tab:abl-asd} reports BD-Distortion in DISTS, FID, and ME relative to the full model~(v), with positive values denoting degradation.
Without DMD, the distortion-only student~(i) is worse than the full model~(v) on all three metrics, showing that source-referenced supervision alone is insufficient for the same overall reconstruction quality.
The qualitative comparison in Fig.~\ref{fig:dmd-visual} further confirms the perceptual benefit of gated DMD.
Together, these results establish the importance of the DMD term within our adaptive objective for recovering realistic fine detail in the one-step codec.

\textbf{Adaptive Score Distillation.}
Variant (iii) directly isolates the gate: retaining both DMD and $\mathcal{L}_{\text{dist}}$ but removing the gate raises BD-ME by 0.174 on HEVC Class~B and 0.140 on UVG.
The resulting motion stalling is corroborated by the optical flow in Fig.~\ref{fig:4_2}, the training dynamics in Fig.~\ref{fig:4_3}, and the jagged temporal trajectory in Fig.~\ref{fig:2_1}.
When the source-referenced loss is also removed, variant~(ii) performs worst on every metric because no ground-truth-anchored term remains to correct misleading DMD updates.
Conversely, gated DMD without the distortion loss (iv) keeps ME close to the full model but degrades DISTS on both datasets.
This confirms the complementary roles established in Sec.~\ref{sec:pixvdc1-asd}: the gate attenuates poorly aligned DMD feedback, whereas the ground-truth-anchored loss supplies the corrective pull.
Only their combination (v) achieves the best joint DISTS, FID, and ME, validating the effectiveness of Adaptive Score Distillation in stabilizing motion while retaining the perceptual benefit.

\begin{table}[t]
\centering
\setlength{\tabcolsep}{4pt}
\caption{BD-Distortion relative to the full model (v), evaluated over the common rate interval. Positive values denote degradation; lower is better.}
\label{tab:abl-asd}
\resizebox{\columnwidth}{!}{%
\begin{tabular}{lccc|cccccc}
\toprule
 & & & & \multicolumn{3}{c}{HEVC Class B} & \multicolumn{3}{c}{UVG} \\
 & $\mathcal{L}_{\text{dist}}$ & $\mathcal{L}_{\text{DMD}}$ & Gate & {\scriptsize BD-DISTS$\downarrow$} & {\scriptsize BD-FID$\downarrow$} & {\scriptsize BD-ME$\downarrow$} & {\scriptsize BD-DISTS$\downarrow$} & {\scriptsize BD-FID$\downarrow$} & {\scriptsize BD-ME$\downarrow$} \\
\midrule
(i) & \checkmark & $\times$ & $\times$ & 0.007 & 2.49 & 0.134 & 0.003 & 1.62 & 0.033 \\
(ii) & $\times$ & \checkmark & $\times$ & 0.031 & 12.4 & 0.691 & 0.033 & 5.58 & 0.552 \\
(iii) & \checkmark & \checkmark & $\times$ & 0.003 & 1.53 & 0.174 & 0.002 & 3.70 & 0.140 \\
(iv) & $\times$ & \checkmark & \checkmark & 0.010 & 2.25 & 0.017 & 0.006 & 0.06 & 0.018 \\
(v) & \checkmark & \checkmark & \checkmark & \textbf{0.000} & \textbf{0.00} & \textbf{0.000} & \textbf{0.000} & \textbf{0.00} & \textbf{0.000} \\
\bottomrule
\addlinespace[2pt]
\multicolumn{10}{l}{\textcolor{gray}{Anchor: full model~(v).}}
\end{tabular}%
}
\vspace{-3mm}
\end{table}

\section{Conclusion}
\label{sec:conclusion}

We presented GenVC, built on a video diffusion model trained from scratch and jointly optimized with a video codec.
Its global-to-local pixel-space hierarchy yields a strong standalone multi-step generative codec, whose diffusion model subsequently serves as the teacher for one-step distillation.
During distillation, we identified a teacher-side guidance failure in which student-induced inputs can make the frozen teacher provide misleading DMD updates that reinforce motion stalling, and introduced Adaptive Score Distillation to attenuate such updates while retaining the corrective pull of ground-truth-anchored losses.
Experiments show that GenVC achieves SOTA ultra-low-bitrate perceptual quality, with average bitrate savings of 62.5\% at matched LPIPS and 71.3\% at matched FID over GLVC.
Its 478.0M-parameter diffusion model reaches a decoding throughput of 15.1~fps for 1080p on an A100 GPU, significantly surpassing existing diffusion-based generative codecs.

{
    \small
    \bibliographystyle{IEEEtran}
    \bibliography{main}
}


 




\vfill

\end{document}